\documentclass[fleqn,twoside]{article}
\usepackage{espcrc2}

\usepackage{graphicx}
\usepackage[figuresright]{rotating}

\newcommand{\AmS}{{\protect\the\textfont2
  A\kern-.1667em\lower.5ex\hbox{M}\kern-.125emS}}

\title{Chiral Loop Effects in the Quenched 
       Scalar Isovector Meson Propagator 
        \thanks{Talk presented by H. Thacker}}
\author{
        W. Bardeen\address[MCSD]{Fermilab, P.O. Box 500, Batavia, IL 60510},
        A. Duncan\address{Dept. of Physics and Astronomy, University of Pittsburgh,
         Pittsburgh, PA 15260, US},
        and
        E. Eichten\addressmark[MCSD],
        N. Isgur\address{Jefferson Lab, Newport News, VA 23606},
        and
        H. Thacker\address{Department of Physics,
        University of Virginia, 
        Charlottesville, VA 22904}%
       }
       
\begin{document}

\begin{abstract}
The scalar isovector meson propagator is studied in quenched QCD. For the lightest
quark masses used, this propagator is dominated by a quenched chiral loop
effect associated with the $\eta'$-$\pi$ two-meson intermediate state. Both the
time dependence and the pion mass dependence of the effect are well-described
by quenched chiral perturbation theory.
\vspace{1pc}
\end{abstract}

\maketitle

\section{Introduction}

Recent calculations in quenched QCD at light valence quark mass have been refined to
sufficient accuracy to allow detailed comparison with the predictions of quenched
chiral perturbation theory \cite{chlogs,scprop}. The identification of quenched chiral
loop effects in lattice data serves several purposes. It allows one to perform sensible
chiral extrapolations of quenched data and to extract the
parameters of the QCD chiral Lagrangian in the quenched approximation. 
It also provides a standard of comparison for future studies of chiral behavior 
in full and partially quenched QCD, which will be crucial for obtaining complete
control over chiral extrapolations in full QCD. In particular, the study of quenched chiral loop (QCL)
effects has identified several quantities for which the effects of quenching
are clearly observable in the lattice data and in detailed agreeement with
quenched chiral perturbation theory. These quantities should be particularly useful in
studying the transition from quenched to full QCD via partially quenched calculations.
The most striking of the observed QCL effects 
appears in the scalar, isovector (valence or ``connected'') 
meson propagator which I will discuss in this talk.

The spectroscopy of scalar mesons is somewhat complicated phenomenologically. In a 
nonrelativistic quark-model framework, the lowest lying scalars are p-wave triplet states and would be 
expected to be substantially heavier than the lowest vector and pseudoscalar mesons,
which are s-wave. However, the experimental situation is somewhat unclear. There is
an $a_0$ state at 980 MeV, but this is very close to the $K\bar{K}$ threshold, and
may be a 4-quark state ($K\bar{K}$ ``molecule''). The next $a_0$ meson is at 1474 MeV.
There is also a strange scalar $K_0^*$ at 1412 MeV.
Our data for the scalar isovector propagator exhibits not only the negative $\eta'$-$\pi$ QCL
effect, but also clear evidence of a positive short-range component corresponding to
a heavy $a_0$ state. Fitting the data to a chiral Lagrangian which includes a heavy
scalar field, as discussed in Section 3, we obtain a chirally extrapolated
scalar meson mass of $M_{a0}=1350\pm 80$ MeV
(statistical errors only). Further studies are required to estimate lattice-spacing and 
quenching effects, but the result appears to favor the 4-quark interpretation of
the $a_0(980)$, with the states in the 1400-1500 MeV range being the lowest-lying
scalar $q\bar{q}$ states.

\section{Lattice results and one-loop $\chi PT$}

\begin{figure}
\vspace*{3.6cm}
\includegraphics{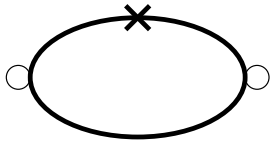}
\caption[]{The one-loop quenched $\chi PT$ contribution to the scalar correlator.}
\end{figure}

While most other QCL effects amount to relatively small modifications of full QCD
behavior even at the lightest quark masses studied, the effect of quenching on the
scalar isovector propagator is striking. This effect arises from the $\eta'$-$\pi$
intermediate state via the chiral loop
graph depicted in Fig. 1. Not only does this graph dominate the entire 
propagator at small quark mass and large time separations, 
but the overall sign of the effect is
{\it negative} i.e. opposite to that required by spectral positivity. (This arises
from a two ghost-meson state in the ghost-quark formulation of quenched $\chi PT$.)
Most of the results presented here were obtained from an ensemble of 300 gauge
configurations on a $12^3\times 24$ lattice at $\beta=5.7$, using clover-improved
fermions with $C_{sw}=1.57$. The MQA pole-shifting ansatz was used to resolve 
the exceptional configuration problem. Nine quark masses were
used, corresponding to pion masses ranging from about 280 $MeV$ to 710 $MeV$. 

\begin{figure}
\vspace*{3.6cm}
\includegraphics{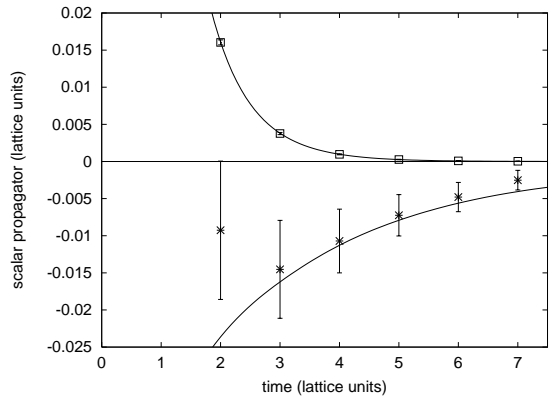}
\vspace{1.5cm}
\caption[]{The scalar isovector correlator for $\kappa=.1400$ (boxes) and $\kappa=.1428$
($\times$'s). The solid curves are an exponential fit (upper curve) and the one-loop
quenched $\chi PT$ estimate of the $\eta'$-$\pi$ state contribution (lower curve),
given by Eqns. (1) and (2).
.}
\end{figure}

The QCL effect in the scalar isovector channel is clearly visible in the bizarre mass
dependence of the raw lattice propagator. In Fig. 2 we show this propagator for 
the heaviest and lightest quark masses studied. For the heavier quarks, the correlator is
positive and exponentially falling, as would be expected for a heavy scalar meson
state of mass $M_{a0}=1.4$-$1.5$ GeV. The correlator completely changes 
character as the quark mass becomes small, developing a negative long-range component.
This component is convincingly explained as the effect of the $\eta'$-$\pi$ chiral
loop diagram in Fig. 1. The solid curve in Fig. 2 is the {\it zero parameter} prediction
of one-loop chiral perturbation theory, which uses only parameters obtained from previous
chiral analysis of the pseudoscalar propagator. Define the one-loop bubble integral,
Fig. 1, in momentum space
\begin{equation}
B(p) = \frac{1}{VT}\sum_q\frac{1}{((q+p)^2+m_{\pi}^2)}\frac{m_0^2}{(q^2+m_{\pi}^2)}
\end{equation}
where $m_0$ is the $\eta'$ mass insertion, which has been determined \cite{chlogs} by
studying the pseudoscalar hairpin correlator ($m_0 = 0.33(2)$ in lattice units at $\beta=5.7$). 
The quenched scalar propagator is dominated by the $\eta'$-$\pi$ intermediated state
and given by
\begin{equation}
\Delta(p) \approx -C_B^2B(p)
\end{equation}
where $C_B$ is the $\bar{\psi}\psi$ to $\eta'\pi$ coupling constant. In a chiral Lagrangian
framework, this coupling is not independently adjustable, but is determined by the 
slope parameter\cite{chlogs}
\begin{equation}
r_0 = m_{\pi}^2/2m_q \approx f_P/f_A
\end{equation}
The constant $C_B$ is related to $r_0$ by a soft pion theorem:
\begin{equation}
C_B = 2r_0
\end{equation}
The excellent agreement between the long-range correlator and the $\chi PT$ prediction exhibited
in Fig. 2 shows that this soft pion theorem is quite well satisfied. 
It is interesting to note that, with $C_B$ fixed by current algebra,
the overall size of the one-loop QCL effect in the scalar propagator
is determined by the $\eta'$ mass insertion $m_0^2$. If we take this as a fit parameter,
we obtain $m_0=0.34(4)$. This is nearly as good a determination of the $\eta'$ mass parameter as the
one obtained directly from the hairpin correlator.

\section{Chiral Lagrangian analysis}

A full chiral Lagrangian description of the lattice data for the scalar isovector propagator
requires the incorporation of a heavy scalar field into the Lagrangian. The formalism of
nonlinear chiral Lagrangians \cite{Weinberg} allows the incorporation of heavy fields which
only have specified transformation properties under the unbroken vector subgroup of the chiral
group. To first order in symmetry breaking and ignoring derivative coupling 
terms, the relevant chiral Lagrangian terms are
\begin{eqnarray}
L & = &\frac{f^2}{4}tr\left(\partial U\partial U^{\dag}\right) + \frac{f^2}{4}tr\left(
\chi^{\dag}U + h.c\right) \nonumber\\
& + &\frac{1}{4}tr\left(D\sigma D\sigma\right)
-\frac{1}{4}m_s^2 tr\left(\sigma\sigma\right) \nonumber\\
& + &f_s tr\left(\chi^{\dag}\sqrt{U}
\sigma\sqrt{U} + h.c.\right) + L_{hairpin}
\end{eqnarray}
 where the last term is the anomaly-induced hairpin mass insertion,
\begin{equation}
L_{hairpin} = -\frac{i}{2}m_0^2(f^2/8)\left[tr\ln(U^{\dag})- tr\ln(U)\right]^2
\end{equation}
The values of $f$ and $m_0$ have been determined by analysis of the pseudoscalar valence and
hairpin correlators.
We are interested in the correlator for a flavor nonsinglet quark bilinear $\bar{\psi}_1\psi_2$.
This operator is related to the chiral Lagrangian fields via the dependence of L on the
spurion field $\chi$. This implies that the quark bilinear consists of two terms, which 
describe its coupling to the $\eta'$-$\pi$ and scalar meson states, respectively,
\begin{eqnarray}
\bar{\psi}_2\psi_1 & = &-\frac{1}{2}r_0f^2\left(U+U^{\dag}\right)_{12}\\
& - & r_0f_s\left(\sqrt{U}\sigma \sqrt{U} + h.c.\right)_{12}
\end{eqnarray}

Using this chiral Lagrangian analysis, a resummed-bubble expression for the $\bar{\psi}_2\psi_1$
correlator was derived and used to fit the lattice correlator \cite{scprop}. The scalar meson
parameters obtained for $\beta=5.7$ are, in lattice units, in the chiral limit
\begin{eqnarray}
f_s & = & .057(3) \\
m_s & = & 1.14(7)
\end{eqnarray}
Fits of the data to the resummed bubble expression for several quark masses are shown in Fig. 3.
It is particularly significant that, over the entire range of quark masses, the fit parameters
$f_s$ and $m_s$ show very little mass dependence, indicating that the dramatic mass dependence
of the data is almost entirely explained by the change of the pion mass in the chiral loop integral.

\begin{figure}
\vspace*{3.6cm}
\includegraphics{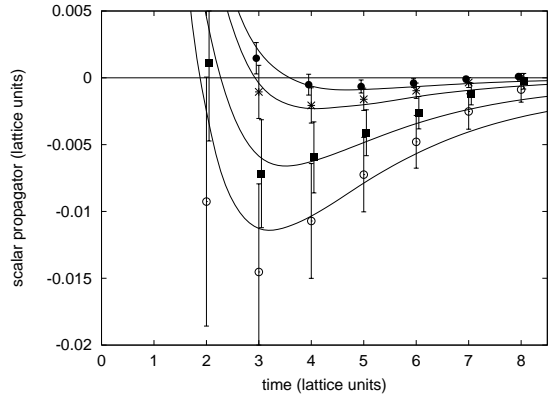}
\vspace{1.5cm}
\caption[]{Scalar isovector correlator for $\kappa=$.1423, .1425, .1427, and .1428.
Solid curves are fits to resummed bubble expression obtained from the chiral Lagrangian (5).}
\end{figure}

\end{document}